# Nanofocusing of Hyperbolic Phonon-Polaritons in a Tapered Boron Nitride Slab

*Alexey Yu. Nikitin[1,2\*†], Edward Yoxall[1†], Martin Schnell[1,3], Saül Vélez[1], Irene Dolado[1], Pablo Alonso-Gonzalez[1,4], Felix Casanova[1,2], Luis E. Hueso[1,2] and Rainer Hillenbrand[1,5\*]*

[1]CIC nanoGUNE, 20018, Donostia-San Sebastián, Spain.

[2]IKERBASQUE, Basque Foundation for Science, 48011 Bilbao, Spain.

[3]Beckman Institute for Advanced Science and Technology, University of Illinois, 61801, Urbana-Champaign, Illinois, USA

[4]Departamento de Física, Universidad de Oviedo, 33007 Oviedo, Spain

[5]CIC NanoGUNE and UPV/EHU, 20018, Donostia-San Sebastian, Spain

**Abstract**

**Nanofocusing of light offers new technological opportunities for the delivery and manipulation of electromagnetic fields at sub-diffraction limited length scales. Here, we show that hyperbolic phonon-polariton (HPP) modes in the mid-infrared as supported by a hexagonal boron nitride (h-BN) slab can be nanofocused (i.e. both field enhanced and wavelength compressed) by propagation along a vertical taper. Via numerical simulations, we demonstrate that field enhancement factors of 90 – for steep tapers – and wavelength compression of more than an order of magnitude – for adiabatic tapers – can be expected. Employing scattering-type scanning near-field optical microscopy (s-SNOM), we provide – for the first time – proof-of-principle experimental evidence of a significant HPP wavelength compression. We expect these functionalities to provide diverse applications, from biosensing and non-linear optics to optical circuitry.**

**Keywords**: nanofocusing, hyperbolic polaritons, wavelength compression, tapered waveguide, BN

The principle objective of nanooptics is to control the flow of light at the nanoscale. It can be achieved through polaritons - electromagnetic modes formed by coupling light with quasi-particles such as plasmons, phonons or excitons – which allow for confining and enhancing electromagnetic fields within nanoscale volumes[1]. Polaritons thus form the basis for a new generation of optical devices and applications[2-4], for example single molecule sensing[5, 6], optical trapping[7], nanophotonic circuitry[8, 9], nonlinear optics[10], coupling between polaritons[11] and slow light[12].

The combination of nanoscale field confinement and field enhancement is generally known as nanofocusing, and has been widely examined in various geometries and at various frequencies between the visible and microwave range[13]. It can be achieved by different mechanisms, the most prominent being localized antenna resonances[14] or compression of surface polariton propagating along tapered slabs, rods and other waveguiding structures[12, 15-22].

Here, we study a method of nanofocusing hyperbolic phonon-polaritons (HPPs) by propagation along a tapered hexagonal boron nitride (h-BN) slab at mid-infrared frequencies. HPPs occur in materials with hyperbolic dispersion[23, 24], where one of the principle components (axial component, indexed $z$) of the dielectric tensor $\hat{\epsilon}$ is opposite in sign to the other two (transverse components, indexed $x$ and $y$). Electromagnetic waves inside hyperbolic materials – the HPPs themselves – thus propagate with a wavevector $\boldsymbol{k}$, which must satisfy the following dispersion relation (where $\epsilon_x = \epsilon_y$):

$$k_z^2/\epsilon_x + \left(k_x^2 + k_y^2\right)/\epsilon_z = (\omega/c)^2. \qquad (1)$$

The solutions of Eq. 1 are open hyperboloids of type I ($\text{Re}(\epsilon_z) < 0, \text{Re}(\epsilon_x) > 0$) or type II ($\text{Re}(\epsilon_z) > 0, \text{Re}(\epsilon_x) < 0$), and they allow extremely large $\boldsymbol{k}$-values to exist. Another consequence of the hyperbolic dispersion is that light within a hyperbolic material travels as well-defined rays, as the propagation direction is fixed within certain cones[23]. As such, HPPs have already been used in diverse applications such as hyperlensing[25] and the control of spontaneous emission[26, 27].

H-BN is a naturally occurring layered crystal[28], which exhibits hyperbolic dispersion in the spectral regions between 760 – 825 $cm^{-1}$ (type I) and 1370 – 1610 $cm^{-1}$ (type II). Because of the sharp h-BN phonon resonances, HPPs in h-BN slabs have longer propagation lengths than, for instance, surface phonon polaritons in amorphous oxide films (e.g. $SiO_2$ or $Al_2O_3$) or surface plasmon polaritons in semiconductors[4]. Importantly, h-BN films can be fabricated by exfoliation – much easier than thin films of high-quality SiC crystals in which phonon polaritons have sharp resonances[29] comparable to that of h-BN. In the following, we focus exclusively on the type II region. HPPs supported by h-BN have been the subject of considerable study recently[30-32]: benefitting as they do from strong field confinement and low losses, they are promising candidates for tailored thermal emission[33], flatland optics[34], enhanced molecular spectroscopy and super-resolution imaging[35, 36].

It has been demonstrated that by tapering hyperbolic metamaterial slabs it is possible to improve the coupling of HPPs to free-space radiation, owing to the dependence of the HPP wavelength on the metamaterial thickness[27]. Here we demonstrate by numerical calculations both the electric field enhancement and wavelength compression – together amounting to nanofocusing – of HPPs when they propagate along a tapered slab of h-BN (illustrated in Fig. 1). Via scanning near-field optical microscopy (s-SNOM) we visualize in real space the HPP propagating into an h-BN taper, providing direct experimental evidence the wavelength compression. A gold edge acts like a broadband antenna, converting the incident radiation into highly confined near fields, which have sufficiently high *k*-values to meet the momentum matching condition and launch HPPs. As these HPPs propagate within a thin slab of h-BN, their rays reflect from the top and bottom surfaces, forming a zig-zag style ray (red zig-zag line in Fig. 1), which can be described by a superposition of the slab's eigenmodes[30, 31, 35-37] Mn, with n = 0,1,2… (the field profiles of the modes M0 and M1 are shown in Fig. 1). These slab eigenmodes do not have a cutoff at large wavevectors, and enable extreme subwavelength confinement of electromagnetic fields. Because the HPPs are volume modes (i.e. the energy flows to a large amount inside the h-BN slab), they are expected to be much less sensitive to scattering by surface defects than surface polaritons. Each eigenmode has its own wavevector $K_x = k_x + i\gamma$. The calculated dispersion curves for the first five modes M0 – M4, (see Methods), are shown in Fig. 2a for a slab of thickness $d$ = 150 nm. Note that the higher order modes do not propagate as far as M0, essentially due to their much shorter wavelengths. Most importantly, the mode dispersion substantially depends on the thickness, $d$[31]. In Fig. 2b, therefore, we show the dispersion of the fundamental mode M0 (with the longest propagation length) for different thicknesses $d$: as $d$ decreases, $k_x$ increases for any given frequency $\omega$. This increased confinement for thinner slabs is clearly visible in Fig. 2c, where the calculated $x$-component of the M0 mode's field distribution, $E_x$, at a frequency $\omega$ = 1500 cm$^{-1}$, (see Methods), is shown for $d$ = 100 nm (upper panel) and $d$ = 50 nm (lower panel). Consequently, when the M0 mode propagates along a tapered slab, we expect an increasing field confinement coming along with an increasing field enhancement (assuming a nearly adiabatic process, i.e. weak dissipation and backreflection of the mode), which is the basis for the HPP nanofocusing method described in this paper.

In order to demonstrate the propagation and nanofocusing of HPPs in tapered h-BN slabs quantitatively, we have performed full-wave numeric simulations (see Methods). We have calculated the electromagnetic fields inside the slab, generated by the mode M0 entering into the tapered part of the slab from the region of the constant thickness of 150 nm. Fig. 3a quantifies the field enhancement at the end of the taper as a function of the taper angle $\theta$. The field enhancement is defined as the ratio between the magnitude of the field on the h-BN surface at the start of the taper (labelled A in the lower panel of Fig. 3b) and close to the end of the taper, where the slab has a thickness of 3 nm (labelled B). We chose this value of 3 nm because in real samples the edges are not ideally sharp, and therefore have a finite thickness at the end of the taper (which defines the maximal field enhancement). Note that the thickness of exfoliated h-BN flakes has reached a few monolayers[38] and thus the presented field enhancement is potentially achievable experimentally. In Fig. 3a, we plot the field enhancement as a function of the taper angle $\theta$. For shallow angles ($\theta$ <

10°), the field enhancement steadily increases with larger $\theta$. In this regime (when the change of the thickness variation along the taper is slow compared with the wavelength of the M0 mode), the HPP is adiabatically compressed: it does not "feel" the slab taper, and its characteristics are simply determined by the local structure's parameters. The enhancement, however, is relatively small (less than 30), as the comparatively large taper lengths at these angles allow for significant losses. The adiabatic compression of the M0 mode is seen in Fig. 3b, where the real part (upper panel) and magnitude (lower panel) of $E_z$ for a slab with $\theta = 3°$ is displayed. It is clear that, while the HPP wavelength is indeed strongly compressed (by a factor of 50), there is no significant enhancement in the near-field distribution. Note that for a taper angle of $\theta = 8°$ we find both adiabatic wavelength compression and field enhancement at the taper end (Fig. S3, Supporting Information).

At steeper taper angles ($\theta > 10°$), however, the adiabatic approximation no longer applies, and there is significant reflection and scattering of the HPP M0 mode at the taper (see Fig.S1 in the Supporting Information). Particularly, at the beginning of the taper (exhibiting a sharp edge marked by X in the lower panel of Fig. 3c), the incoming HPP M0 mode scatters into higher-order HPP modes. The interference of the latter yields a complex pattern of HPP rays[37], which are clearly visible in the near-field maps of Fig. 3c. Interestingly, the HPP rays do not appear in the adiabatic regime (Fig. 3b), where the edge at the beginning of the taper is not sharp enough to efficiently scatter the incoming M0 HPP mode. We stress that the scattering of the M0 HPP mode into higher-order modes has a large impact on the field enhancement at the taper end: due to interference of the generated HPP rays, the field enhancement at the taper end oscillates when the taper angle is increased (Fig. 3a). The field enhancement factor peaks at 92 for $\theta = 27.5°$, and this is clearly visible as a "hot spot" in both panels of Fig. 3c, while the wavelength compression is no longer easily seen due to the small length of the taper compared to the HPP wavelength. Such large field enhancements compare favourably with that provided by infrared metallic antennas[39].

Finally, we present a proof-of-principle experiment, where s-SNOM was used to map the near fields above a tapered slab of h-BN (illustrated in Fig. 4b). A p-polarized infrared beam at frequency $\omega = 1500$ cm$^{-1}$ and with electric field $E_{in}$ illuminates both the sample and the silicon probe of our s-SNOM. The sample consists of a tapered h-BN slab onto which we deposited a gold film (Fig. 4a shows a light microscope image), whose edge launches HPPs in the h-BN slab[37] (see Methods for the fabrication of the sample). While the silicon probe tip (oscillating vertically at frequency $\Omega$) is scanned perpendicularly to the gold edge and across the h-BN taper (parallel to the dashed white line in Fig. 4a), the tip-scattered light $E_{scat}$ – a measure of the near field above the h-BN surface – is detected interferometrically and demodulated at frequency $3\Omega$, yielding $E_{scat,3}$ (see Methods). For imaging the HPPs, we recorded $|E_{scat,3}|$ as a function of the tip position $x$. Note that the fabrication of our samples is limited to naturally occurring tapers created by h-BN exfoliation. Therefore, the tapers are typically very shallow (~ 1.5° in our case), so we expected to see an adiabatic wavelength compression of the HPP field rather than a significant field enhancement

The gold edge strongly launches HPPs, whose electric field is labelled as $E_{HPP}$ in Fig. 4b, while the silicon tip launches very weakly[37] (in contrast to previous s-SNOM studies of h-BN HPPs, where a metallic tip is used as the launcher of HPPs[31, 32]). Pure scattering of polaritons by the tip (rather than exciting polaritons with the tip) allows for mapping of the wavefronts of the polaritons[40], thus directly visualizing the wavelength compression when they propagate into a taper. The magnitude of the detected and demodulated tip-scattered light, $|E_{scat,3}|$, averaged over 70 line scans (the path of one of the scans is illustrated by the white dashed line in Fig. 4a), is shown in the upper panel of Fig. 4c by the red curve. Fringes, whose height decays with increasing distance from the gold edge, are clearly discerned. The fringes arise from the alternating constructive and destructive interference between the incident field $E_{in}$ and the HPP field $E_{HPP}$. Strikingly, the spacing between the fringes (indicating the HPP wavelength) reduces with increasing distance from the gold edge, which is in good agreement with a numerical simulation of the magnitude of the total electric field's $z$-component[41], $E_{z,in} + E_{z,HPP}$, (Fig. 4c, lower panel). In this simulation we used the profile of the h-BN flake, averaged over 70 line scans (shown by the continuous white outline in Fig. 4c, lower panel) obtained from the AFM topography imaging. Note that the gold edge launches many HPP modes – hence rays are visible within the h-BN slab seen in the lower panel of Fig. 4c – but only the M0 mode propagates along the full length of the taper.

For quantitative comparison of the s-SNOM signal with our numerical calculation, we have to consider that the demodulated tip-scattered field, $E_{scat,3}$, depends on both the total electric field's $z$-component, $E_{z,in} + E_{z,HPP}$, and the near-field interaction between the tip and the sample. According to the well-established dipole model[42, 43], the s-SNOM signal (in case of a slow vertical decay of the polariton's electric field) is given by $E_{scat,3} \propto \alpha_{eff,3} \cdot (E_{z,in} + E_{z,HPP})$, where $\alpha_{eff,3}$ describes the near-field interaction for signal demodulation at 3Ω. The inset of Fig. 4c (upper panel) shows $\alpha_{eff,3}$ for the h-BN slab as a function of $x$, calculated according to the dipole model for thin layers described in Ref.[43]. For this calculation we used the h-BN thickness obtained from the topography image of the sample, and dielectric values for h-BN from Ref.[30]. The calculated s-SNOM signal $E_{scat,3}$ is shown by the blue curve in the upper panel of Fig. 4c. We find a good agreement with the experiment (red curve) regarding fringe spacing and decay of the fringe maxima with increasing distance $x$ to the gold edge. Particularly, our calculation explains the rather rapid decay of the fringe maxima in the s-SNOM profile (compared to the field calculations in the lower panel of Fig. 4c) due to the strongly decreasing value of $\alpha_{eff,3}$ for decreasing thickness (increasing $x$) of the h-BN slab. Note that the field of the M0 mode strongly decreases upon propagation into the taper. For that reason, the field associated with HPPs back-reflected at the taper end can be neglected.

In principle, the wavelength of the HPP's fundamental mode M0 can only be extracted from the fringe spacing after correcting for the angled illumination (40°): in practice, however, this correction is negligibly small due to the short wavelength of the HPP in relation to the free-space wavelength[37, 42]. The fringe spacing in both the experiment and simulation thus directly yields the HPP's M0 wavelength, $\lambda_{M0}$. In Fig. 4d we show the experimental local wavelength $\lambda_{M0}$ (red circles in Fig. 4d) obtained from the red s-SNOM profile in Fig. 4c and the numerically simulated local wavelength $\lambda_{M0}$ (blue squares in Fig. 4d) obtained from the blue curve in Fig. 4c. In both experiment and simulation we clearly see a continuous decrease of $\lambda_{M0}$ with increasing distance x from the gold edge by more than a factor of 2. For comparison, in Fig. 4c we also show the analytically calculated wavelength $\lambda_{MO}$[31] (green curve in Fig. 4d) for uniform slabs. At each position $x$, this calculation assumes a slab thickness $d(x)$, which we obtained from the AFM height profile shown by the solid white line in the lower panel of Fig. 4c. Like in the experiment and numerical simulation (blue and red symbols in Fig. 4c), we find that the analytically calculated $\lambda_{MO}$ decreases with decreasing slab thickness. Quantitative differences for $x > 2$ μm we explain by the backreflection of the HPP M0 mode at the slab termination ($x = 4.5$ μm), which is not taken into account in the analytical calculation of $\lambda_{MO}$ in uniform slabs. Further, the wavelength $\lambda_{MO}$ continuously decreases along the taper with increasing x. Because of the nonlinear dependence of $\lambda_{MO}$ on $d(x)$, the peak-to-peak distance $\Delta x$ in the simulated blue s-SNOM profile does not provide the exact wavelength $\lambda_{MO}$ at position $x$ but rather a good approximate value. Our experimental results therefore provide the first direct observation of HPP compression upon propagation along a tapered slab.

In conclusion, we have numerically demonstrated both adiabatic nanofocusing of HPP modes by propagation along a tapered slab of h-BN (leading to wavelength compression of the modes), and non-adiabatic nanofocusing (leading to field enhancements of almost two orders of magnitude at the taper end). Furthermore, we have provided the first experimental evidence of HPP wavelength compression by way of a proof-of-principle s-SNOM measurement, and seen excellent qualitative agreement between the experiment and simulation. We anticipate that s-SNOM will be a vital tool in the further characterisation of polariton nanofocusing devices, and future work in this direction will certainly involve both fabrication and measurement of an h-BN slab with a large taper angle, for which the strong field enhancement predicted by this paper should be experimentally visible. We expect that HPP field enhancement will find applications in photodetection, sensing and non-linear optics, while the HPP wavelength compression provides a potential avenue for coupling different types of polaritons, as may be required by future optical circuitry designs.

**Methods**

*Fabrication of BN films with gold stripes*

The h-BN slab was prepared by mechanical exfoliation of commercially available h-BN crystals (HQgraphene Co, N2A1) using blue Nitto tape (Nitto Denko Co., SPV 224P) and transferred onto a Si/SiO$_2$(250nm) substrate. Following this, a wide gold stripe (to launch the HPPs) was deposited by electron beam lithography along a naturally tapered h-BN edge with the help of an electro-sensitive photoresist (PMMA), high resolution electron beam lithography, metal deposition (3nm-thick Ti stick layer deposited via e-beam evaporation followed by thermal evaporation of a 40nm-thick Au film) and lift-off.

*Near-field microscopy*

Our s-SNOM (Neaspec, Munich) is based on an atomic force microscope (AFM). A silicon tip - oscillating vertically at the cantilevers mechanical resonance frequency - acts as a scattering near-field probe. The oscillation amplitude was about 80 nm. p-polarised infrared light from a quantum cascade laser (Daylight solutions) was focused via a parabolic mirror onto both the tip and sample at an angle of 40 degrees with respect to the surface normal. The tip-scattered light of field $E_{scat}$ was recorded with a pseudo-heterodyne interferometer[44]. To suppress background scattering from the tip shaft and the sample, the detector signal is typically demodulated at a higher harmonic *n* of the oscillation frequency, yielding the signal $E_{scat,n}$. In this work, we performed third-harmonic demodulation, n= 3.

We conducted similar experiments with metallic tips. We found that both dielectric and metallic tips launch HPPs much less efficiently than the gold edge. Moreover, the tip-launched HPPs can propagate below the gold film (as they are volume modes), and thus do not exhibit a significant back-reflection at the gold edge. We also did not observe a significant back-reflection of tip-launched HPPs from the taper end, which we explain by the strong damping of the HPPs before they reach the very end of the taper.

*Simulations*

The profiles in Fig. 1 and the dispersion curves in Fig. 2a,b were calculated by finite-finite-difference frequency-domain (FDFD) method, using the commercial software package Lumerical Solutions. The numerical calculations shown in Fig. 2c, Fig. 3 and lower panel of Fig. 4c were done by the finite boundary elements method, using the commercial software package Comsol. The dependency of the wavelength, $\lambda_{MO}$, on the distance along the taper, $x$, shown in Fig. 4d, was calculated according to the approximate analytical solution[31], using the profile of the tapered h-BN slab, $d(x)$, extracted from the AFM measurements. The convergence of the numerical simulations was verified for each case. As an example, the convergence of the field enhancement shown in Fig. 3a is illustrated in Fig. S2 of the Supporting Information.

**Author information**

Corresponding Authors:

*E-mails: r.hillenbrand@nanogune.eu, a.nikitin@nanogune.eu

[†]These authors contributed equally

**Acknowledgements**

The authors acknowledge support from the European Union through ERC starting grants (TERATOMO grant no. 258461, SPINTROS grant no. 257654 and CarbonLight grant no. 307806), the European Commission under the Graphene Flagship (contract no. CNECTICT-604391), the Spanish Ministry of Economy and Competitiveness (national projects MAT2014-53432-C5-4-R, MAT2012-36580, MAT2015-65525-R and MAT2012-37638). M.S. acknowledges funding from the European Union's Horizon 2020 research and innovation programme under the Marie Sklodowska-Curie grant agreement No. 655888 (SYNTOH).

**Competing financial interests**

R.H. is co-founder of Neaspec GmbH, a company producing scattering-type scanning near-field optical microscope systems such as the one used in this study. All other authors declare no competing financial interests.

**ASSOCIATED CONTENT**

**Supporting Information**

The Supporting Information includes additional simulations for the reflection, transmission and out-of-plane scattering of the fundamental HPP mode propagating along the tapered h-BN slab; the convergence of the numeric simulations for the field enhancement; and the nanofocusing of HPPs for the taper angle of 8°. This material is available free of charge via the Internet at http://pubs.acs.org.

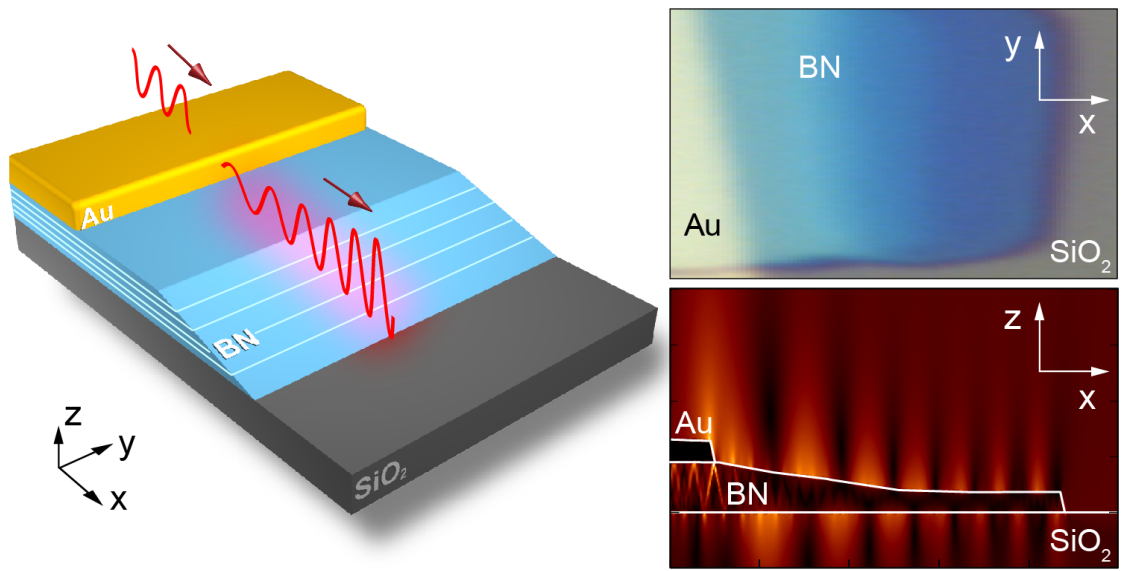

**Figure TOC**

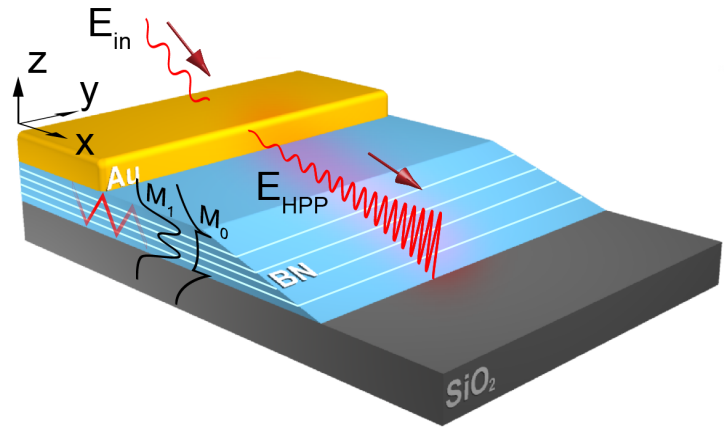

**Figure 1** – Schematics of the HPP compression experiment. The light blue lines of the h-BN slab represent the layers of the material. The red zigzag line illustrates the polaritonic ray launched by the gold edge. The black curves M0 and M1 show the field profiles of the first two modes composing the polaritonic ray.

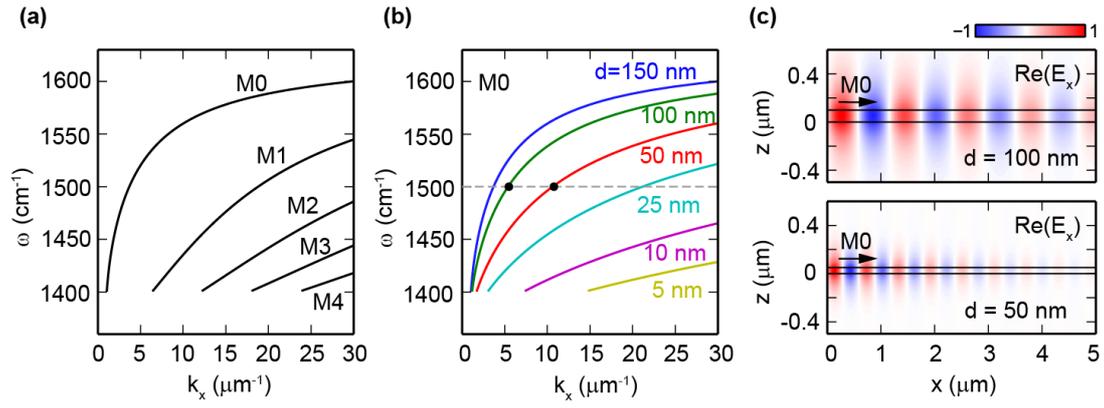

**Figure 2** – (a) Calculated dispersion curves for the first five HPP modes supported by an h-BN slab of thickness $d = 150$ nm. (b) The dependence of the fundamental (M0) mode on the thickness of the slab $d$. The black dots represent the frequency and thicknesses shown in (c). (c) M0 mode profiles showing the *x*-component of the HPP field $E_x$ at frequency $\omega = 1500$ cm$^{-1}$ for $d = 100$ nm (upper panel) and 50 nm (lower panel).

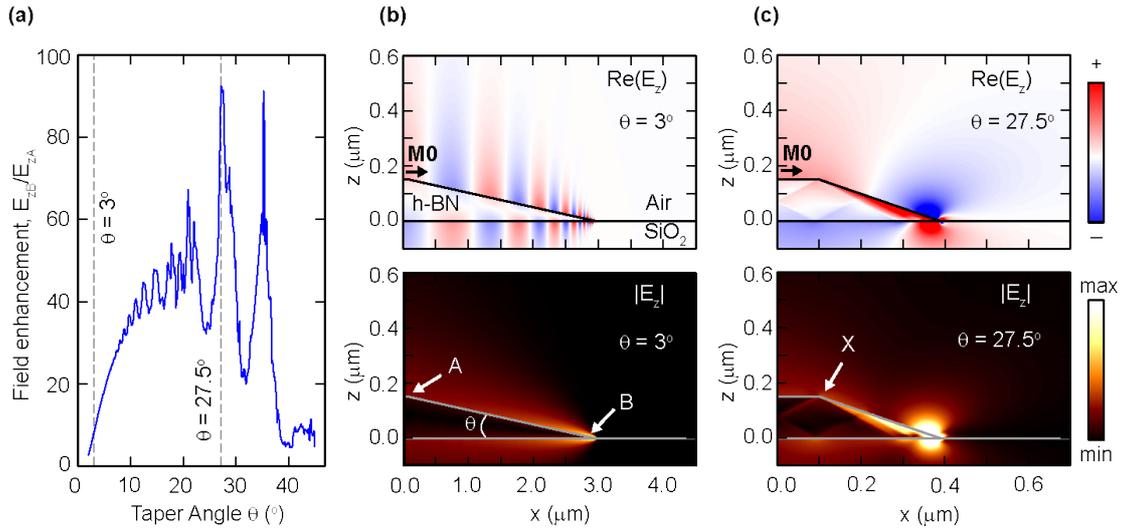

**Figure 3** – (a) Field enhancement as a function of the h-BN taper angle $\theta$. Field enhancement is defined as the ratio between the field near the end of the taper B where the slab has a thickness of 3 nm (labelled B in (b)), and the start of the taper (labelled A). The dashed grey lines mark the values of $\theta$ simulated in (b) and (c). (b) Simulation showing the real part (upper) and magnitude (lower) of the HPP field $E_z$ for a taper angle $\theta$ of 3°, where $\omega = 1500$ cm$^{-1}$. Only the M0 mode is injected into the taper. (c) As for (b), but where $\theta = 27.5°$. The point X marks the beginning of the taper, where higher order HPP modes are generated. Note the different $x$-axis scales between (b) and (c).

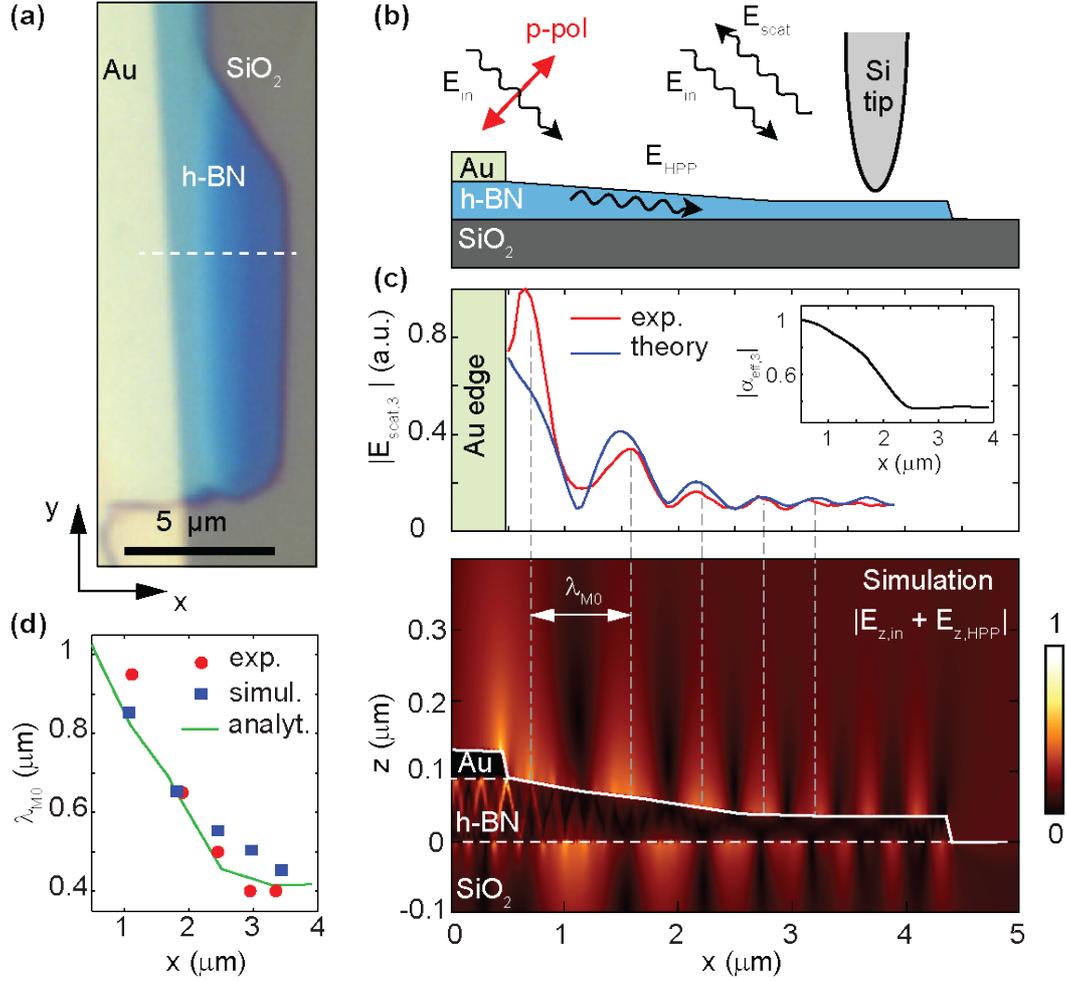

**Figure 4** – (a) Optical microscope image of the studied tapered h-BN flake. (b) Illustration of s-SNOM detection of the HPP compression in a tapered h-BN slab. (c) Upper panel: Red curve shows the experimental s-SNOM profile perpendicularly to the gold edge. The blue curve shows the simulated s-SNOM signal, given by $E_{scat,3} \propto \alpha_{eff,3} (E_{z,in} + E_{z,HPP})$. The inset shows the absolute value of $\alpha_{eff,3}$ as a function of $x$. Lower panel: Numerical calculation of the z-component of the total electric field, $E_{z,in} + E_{z,HPP}$, taking into account the geometry of the h-BN slab (obtained from topography measurements and outlined by the white solid line). (d) Experimental (red circles), calculated according to the field profile (blue squares) and calculated analytically (green line) wavelength of the M0 HPP mode, $\lambda_{M0}$, as a function of distance $x$ along the taper.